\documentclass[aps,prl,showkeys,superscriptaddress, twocolumn]{revtex4}
\usepackage{amsmath}
\usepackage{amssymb}
\usepackage{graphicx}

\renewcommand{\exp}[1]{\ensuremath{{\mathrm e}^{#1}}}
\renewcommand{\kappa}{\varkappa}

\begin{document}

\title{Artificial ferroelectricity due to anomalous Hall effect in magnetic tunnel junctions}

\author{A. Vedyayev}
\email{vedy@magn.ru}
\affiliation{Department of Physics, Moscow Lomonosov State University, Moscow 119991, Russia}
\affiliation{SPINTEC, UMR 8191 CEA-INAC/CNRS/UJF-Grenoble 1/Grenoble-INP, 38054 Grenoble, France}

\author{N. Ryzhanova}
\affiliation{Department of Physics, Moscow Lomonosov State University, Moscow 119991, Russia}
\affiliation{SPINTEC, UMR 8191 CEA-INAC/CNRS/UJF-Grenoble 1/Grenoble-INP, 38054 Grenoble, France}

\author{N. Strelkov}
\affiliation{Department of Physics, Moscow Lomonosov State University, Moscow 119991, Russia}
\affiliation{SPINTEC, UMR 8191 CEA-INAC/CNRS/UJF-Grenoble 1/Grenoble-INP, 38054 Grenoble, France}

\author{B. Dieny}                                                                                           
\email{bernard.dieny@cea.fr}                                                                                
\affiliation{SPINTEC, UMR 8191 CEA-INAC/CNRS/UJF-Grenoble 1/Grenoble-INP, 38054 Grenoble, France}

\begin{abstract}
We theoretically investigated Anomalous Hall Effect (AHE) and Spin Hall Effect (SHE) transversally to the
insulating spacer O, in magnetic tunnel junctions of the form F/O/F where F are ferromagnetic layers
and O represents a tunnel barrier. We considered the case of purely ballistic (quantum mechanical)
transport, taking into account the assymetric scattering due to spin-orbit interaction in the tunnel barrier.
AHE and SHE in the considered case have a surface nature due to proximity effect. Their amplitude is
in first order of the scattering potential. This contrasts with ferromagnetic metals wherein these effect are
in second (side-jump scattering) and third (skew scattering) order on these potentials. The value of AHE
voltage in insulating spacer may be much larger than in metallic ferromagnetic electrodes.
For the antiparallel orientation of the magnetizations in the two F-electrodes, a spontaneous Hall voltage exists even
at zero applied voltage. Therefore an insulating spacer sandwiched between two ferromagnetic layers
can be considered as exhibiting a spontaneous ferroelectricity.
\end{abstract}

\keywords{Hall effect, anomalous Hall effect, spin Hall effect, spintronics, ferroelectrics}

\maketitle

The Anomalous Hall Effect (AHE) in ferromagnetic metals and Spin Hall Effect (SHE) in nonmagnetic materials
have attracted a renewed interest in the last decades. One can notice that AHE and SHE have the same origin,
namely spin-orbit interaction in the presence of magnetic ordering for AHE and without magnetic ordering
for SHE.  Detailed analyses of the mechanisms responsible for these two effects may be found in
reviews~\cite{andp.200510206,JPSJ.75.042001,RevModPhys.82.1539}.
These mechanisms are divided into two groups: intrinsic ones and extrinsic ones. The former appear in pure
metals and have topological nature, closely connected with Berry curvature. Extrinsic mechanisms are due to
asymmetric electron scattering on defects in presence of spin-orbit interaction. Two main types of scattering are
considered: skew scattering~\cite{S0031-89145592596-9,PhysRevB.8.2349} and side-jump
scattering~\cite{PhysRevB.2.4559}. Most of theoretical papers on AHE and
SHE considered the case of infinite homogeneous samples.
References~\cite{PhysRevB.57.2943,0304-88539490469-3} also
investigated AHE for multilayers and for highly inhomogeneous media.

\begin{figure}[ht]
\includegraphics[width=0.45\textwidth]{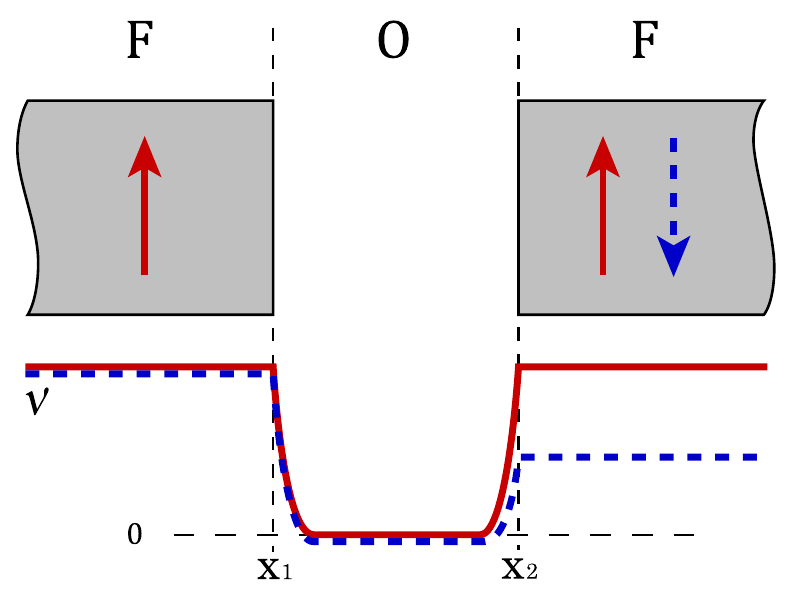}
\caption{\label{fig:sys} Schematic illustration of MTJ. F - ferromagnetic layers, O - insulating spacer.
Arrows denote the direction of magnetizations in electrodes, for parallel (P) and antiparallel (AP) 
orientations. Inset schematically illustrates the dependence of density of states
($\nu$ - in arbitrary units) of spin up tunnelling electrons on the distance from the interface for
P and AP orientations.}
\end{figure}

Let's consider a magnetic tunnel junctions (i.e. a sandwich of two ferromagnetic layers separated by a dielectric
spacer, MTJ -- Magnetic Tunnel Junction) (Fig.\ref{fig:sys}) submitted to a bias voltage applied between
the two F-electrodes supposed to be made of the same ferromagnetic material.
In this study, we are primarily interested by the Hall voltage which may appear between the opposite sides
of the tunnel barrier due to the Hall current inside the spacer in presence of spin-orbit scattering
on impurities. We will show that these Hall and spin Hall current do exist and that moreover, for the
antiparallel orientation of the magnetizations in the two ferromagnetic layers, a spontaneous transverse
Hall voltage exists, even in the absence of any applied bias voltage.

The Hall currents were calculated using Keldysh formalism~\cite{KeldyshTech}. The electrons were described
as forming a free electron gas submitted to $s$-$d$ exchange interaction. As an example, the
Green functions for the considered system (Fig.\ref{fig:sys}) and for $z$-projection of electron's spin
antiparallel to the magnetization in the left electrode are:
\begin{multline}
G^\uparrow_{\mathrm{AP}, x>x'}(r,r')=\frac1N\sum_\kappa
-\frac{1}{2q\mathfrak{D}}\exp{i\kappa_i(y-y')}\exp{i\kappa_z(z-z')}\\
\times\Bigl( \exp{q(x-x_2)}(q+ik_2)+\exp{-q(x-x_2)}(q-ik_2) \Bigr)\\
\times\Bigl( \exp{q(x'-x_1)}(q-ik_1)+\exp{-q(x'-x_2)}(q+ik_1) \Bigr), \label{eq:greenfunc1}
\end{multline}
\begin{multline}
G^\uparrow_{\mathrm{AP}, x<x'}(r,r')=\frac1N\sum_\kappa
-\frac{1}{2q\mathfrak{D}}\exp{i\kappa_i(y-y')}\exp{i\kappa_z(z-z')}\\
\times\Bigl( \exp{q(x-x_1)}(q-ik_1)+\exp{-q(x-x_2)}(q+ik_1) \Bigr) \label{eq:greenfunc2}\\
\times\Bigl( \exp{q(x'-x_2)}(q+ik_2)+\exp{-q(x'-x_2)}(q-ik_2) \Bigr),
\end{multline}
where:
\begin{equation}
\mathfrak{D}=\left(\exp{qb}(q-ik_1)(q-ik_2)-\exp{-qb}(q+ik_1)(q+ik_2)\right), \nonumber
\end{equation}
\begin{equation}
q=\sqrt{\frac{2m}{\hbar^2}(U-E)+\kappa^2},\label{eq:q_k}
\end{equation}
\begin{equation}
k_{1,2}=\sqrt{\frac{2m}{\hbar^2}(E\pm J_{sd})-\kappa^2}.\nonumber
\end{equation}

In (\ref{eq:greenfunc1}) and (\ref{eq:greenfunc2}) ``AP'' means the antiparallel orientation of magnetizations
in the two ferromagnetic electrodes and $x_1$, $x_2$ -- are the ferromagnetic/insulator
interfaces coordinates. In (\ref{eq:q_k}), $U$ is the barrier height, $E$ is the electron energy,
$J_{sd}$ is $s$-$d$ exchange energy. For the opposite direction of spin, all projection changes in
(\ref{eq:greenfunc1}) and (\ref{eq:greenfunc2}) are straightforward.
From (\ref{eq:greenfunc1}) and (\ref{eq:greenfunc2}),
it follows that for the considered system, a finite density of states exists in the energy gap
within the barrier due to proximity effect, which decreases exponentially with the distance from
F/O interfaces (Fig.\ref{fig:sys}). In other words, a quasi-two-dimensional electron gas exists inside the
barrier near the interfaces. Similarly to three dimensional topological
insulator, this electron gas can give birth to charge and spin currents~\cite{RevModPhys.82.3045}.
\begin{figure}[ht]
\includegraphics[width=0.49\textwidth]{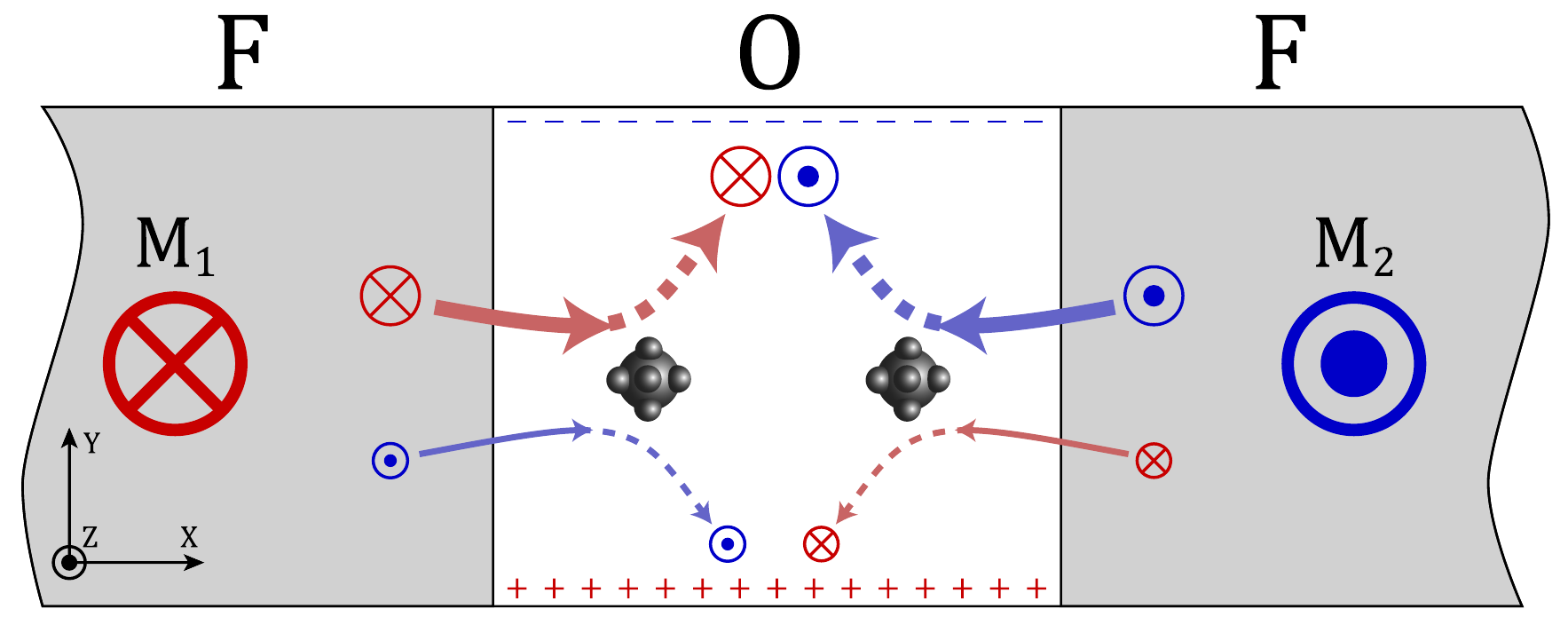}
\caption{\label{fig:sc} Schematic illustration
of AHE and SHE in MTJ due to spin-orbit scattering on impurities. $\otimes$ and $\odot$ denote the
direction of magnetizations and electrons spins.
The thickness of lines are proportional to Hall currents for the given projection of spin.}
\end{figure} 
Evidently the mechanisms of creation of these surface states are different in the two cases.
Let's suppose now that the tunnelling electrons experience scattering on impurities within the barrier with spin-orbit interaction.
This asymmetric scattering deviates the electrons in the direction perpendicular to the tunnel current and to the
projection of their spin. So if the current is spin-polarized, a Hall voltage appears transversally to the tunnel barrier.
Quite interestingly, in antiparallel magnetic configuration of the MTJ, this AHE appears spontaneously even in the
absence of bias voltage applied across the tunnel barrier.
In addition, as illustrated in Fig.\ref{fig:sc}, if the two ferromagnetic materials were assumed to have different
spin-polarization, a spontaneous spin unbalance (spin Hall effect) would also appear between the two
transversal sides of the tunnel barrier at zero bias voltage. This would even be true if one of the ferromagnetic
electrode was replaced by a non-magnetic metallic electrode.

To investigate this effect we added into the free electron Hamiltonian, the impurity potential including
spin-orbit interaction and calculated the induced perturbation to the wave functions:
\begin{multline}
\psi=\psi_0(r)+\int G(r,r')V_{so}(r') \,d^3r'=\\
\psi_0(r)+\int \delta(r'-r_i)(a_0^5\lambda_0)\,d^3r'\label{eq:psi}\\
\times\left[G(r,r')i\sigma_z\left( \frac{\overleftarrow\partial}{\partial x'}
\frac{\overrightarrow\partial}{\partial y'} - \frac{\overleftarrow\partial}{\partial y'}
\frac{\overrightarrow\partial}{\partial x'} \right) \psi_0(r')\right].
\end{multline}

In (\ref{eq:psi}) $\lambda_0$ represents the intensity of spin-orbit interaction, $a_0$ -- lattice parameter,
$r_i$ -- position of the impurity, $\sigma_z$ -- $z$-component of Pauli matrix. Zero order wave
function for the left-to-right and right-to-left tunnelling electrons are correspondently:
\begin{multline}
\psi^\uparrow_{\mathrm{AP}, l}=\frac{2\sqrt{k_1}}{\mathfrak{D}} \\
\times\left( \exp{q(x-x_2)}(q+ik_2)+\exp{-q(x-x_2)}(q-ik_2) \right),\label{eq:psi_l}
\end{multline}
\begin{multline}
\psi^\uparrow_{\mathrm{AP}, r}=\frac{2\sqrt{k_2}}{\mathfrak{D}}\\
\times\left( \exp{q(x-x_1)}(q-ik_1)+\exp{-q(x-x_1)}(q+ik_1) \right).\label{eq:psi_r}
\end{multline}

Now it is easy to calculate the Hall current in ballistic regime in the first order on spin-orbit interaction:
\begin{multline}
\label{eq:jH}
j^\sigma_H=\frac{e}{2\pi\hbar}\int \frac{f(E)}{(2\pi)^2}\,dE\\
\times\int i\sigma_z
\left(
\psi_l^\sigma\frac{\partial}{\partial y}\psi_l^{\sigma *} -
\psi_l^{\sigma *}\frac{\partial}{\partial y}\psi_l^\sigma
 \right)^{(1)}
\,d\kappa_y d\kappa_z\\
+\frac{e}{2\pi\hbar}\int \frac{f(E+eV)}{(2\pi)^2}\,dE\\
\times\int i\sigma_z
\left(
\psi_r^\sigma\frac{\partial}{\partial y}\psi_r^{\sigma *} -
\psi_r^{\sigma *}\frac{\partial}{\partial y}\psi_r^\sigma
 \right)^{(1)}
\,d\kappa_y d\kappa_z,
\end{multline}                                                          
where $f(E)$ -- Fermi distribution for the left electrode, $f(E+eV)$ -- the same for the right one,
$V$ -- applied voltage.
Subscript ``$(\dots)^{(1)}$'' denotes the first order terms on spin-orbit interaction in the
expression in brackets.

Substituting (\ref{eq:psi}), (\ref{eq:psi_l}) and (\ref{eq:psi_r}) into (\ref{eq:jH}) and averaging on the
position of impurities $r_i$ yields the following expressions for the spin-up all current originating
respectively from left ($l$) and right ($r$) electrodes in AP configuration:
\begin{multline}
\label{eq:j_l}
j^\uparrow_{\mathrm{AP}, l}\sim\int d\kappa_y d\kappa_z dE
\frac{4\lambda_0 \kappa_y^2 k_1}{\left|\mathfrak{D}\right|^4}f(E)\\
\times\Biggl[
\left(\exp{2q(x-x_2)}\exp{-2qb}-\exp{-2q(x-x_2)}\exp{2qb}\right)(q^2+k_2^2)^2\\
+2\left(\exp{2qb}-\exp{-2qb}\right)(q^2-k_2^2)(k_1^2-k_2^2)\\
+\left(\exp{2q(x-x_1)}-\exp{-2q(x-x_1)}\right)\\
\times(q^2+k_2^2)\left((q^2+k_1^2)+(k_1^2-k_2^2)\right)\\
-2\left(\exp{2q(x-x_2)}-\exp{-2q(x-x_2)}\right)\\
\times\left((q^2-k_1k_2)^2-q^2(k_1+k_2)^2\right)
\Biggr],
\end{multline}
\begin{multline}
\label{eq:j_r}
j^\uparrow_{\mathrm{AP}, r}\sim\int d\kappa_y d\kappa_z dE
\frac{4\lambda_0 \kappa_y^2 k_2}{\left|\mathfrak{D}\right|^4}f(E+eV)\\
\times\Biggl[
\left(\exp{2q(x-x_1)}\exp{2qb}-\exp{-2q(x-x_1)}\exp{-2qb}\right)(q^2+k_1^2)^2\\
+2\left(\exp{2qb}-\exp{-2qb}\right)(q^2-k_1^2)(k_1^2-k_2^2)\\
+\left(\exp{2q(x-x_2)}-\exp{-2q(x-x_2)}\right)\\
\times(q^2+k_1^2)\left((q^2+k_2^2)-(k_1^2-k_2^2)\right)\\
-2\left(\exp{2q(x-x_1)}-\exp{-2q(x-x_1)}\right)\\
\times\left((q^2-k_1k_2)^2-q^2(k_1+k_2)^2\right)
\Biggr].
\end{multline}

From (\ref{eq:j_l}) and (\ref{eq:j_r}), it follows that the Hall current exponentially depends on the
coordinate $x$ and reaches its maximum near the ``left'' interface for the ``left'' electrons and at
the ``right'' interface for ``right'' electrons. This emphasizes the surface nature of the considered
Hall effect.

It's interesting to note that the present Hall effect in MTJ barrier appears at first order on
the scattering potential, whereas for infinite ferromagnetic metals the Hall effect is in
third order on the scattering potential for skew scattering and in second order for side jump
mechanism~\cite{andp.200510206,JPSJ.75.042001,RevModPhys.82.1539}.
This difference is due to the strong inhomogeneity of the considered system in $x$-direction.
The other remarkable difference already pointed out is that this Hall effect spontaneously exists
even at zero bias voltage in MTJ.

Next the obtained expressions for Hall currents and spin Hall currents were averaged over the
coordinate $x$ and integration over momentum $\vec \kappa$ and energy $E$ yields in the
limit $\exp{-2qb}\ll 1$, in parallel configuration of the MTJ:
\begin{equation}
\label{eq:j_p_h_skew}
\left<j_{l+r}^{\uparrow+\downarrow}\right>_{\mathrm{AHE}}^{\mathrm{P, skew}}=
\frac{4}{15\pi}\,\frac{e^2}{2\pi\hbar}\,\frac{\tilde \lambda c}{U^2b}
\left(E_F^{\uparrow 2}k_F^\uparrow-E_F^{\downarrow 2}k_F^\downarrow\right)V,\nonumber
\end{equation}
\begin{equation}
\label{eq:j_p_sh_skew}
\left<j_{l+r}^{\uparrow+\downarrow}\right>_{\mathrm{SHE}}^{\mathrm{P, skew}}=
\frac{4}{15\pi}\,\frac{e^2}{2\pi\hbar}\,\frac{\tilde \lambda c}{U^2b}
\left(E_F^{\uparrow 2}k_F^\uparrow+E_F^{\downarrow 2}k_F^\downarrow\right)V,\nonumber
\end{equation}
and in antiparallel configuration:
\begin{multline}
\label{eq:j_ap_h_skew}\nonumber
\left<j_{l+r}^{\uparrow+\downarrow}\right>_{\mathrm{AHE}}^{\mathrm{AP, skew}}=
\frac{8}{105\pi}\,\frac{e}{2\pi\hbar}\,\frac{\tilde \lambda c}{U^2b}
\Biggl(
E_F^{\uparrow 3}k_F^\uparrow-E_F^{\downarrow 3}k_F^\downarrow\\
+\left(E_F^\uparrow+eV\right)^3\sqrt{k_F^{\uparrow 2}+eV}
-\left(E_F^\downarrow+eV\right)^3\sqrt{k_F^{\downarrow 2}+eV}
\Biggr),
\end{multline}
\begin{equation}
\label{eq:j_ap_sh_skew}\nonumber
\left<j_{l+r}^{\uparrow+\downarrow}\right>_{\mathrm{SHE}}^{\mathrm{AP, skew}}=
\left<j_{l+r}^{\uparrow+\downarrow}\right>_{\mathrm{SHE}}^{\mathrm{P, skew}},
\end{equation}
where $\tilde\lambda=2ma_0^2\lambda_0/\hbar^2$ -- dimensionless constant of
spin-orbit interaction, $c$ -- atomic concentration of impurities.

One may notice that in contrast to the tunnelling current through the tunnel barrier, the expressions of the
Hall and spin Hall currents do not contain the small parameter $\exp{-2qb}$. Instead, the averaged Hall
voltage decreases inversely proportional to the barrier thickness. Its amplitude is proportional to the small
parameter $\lambda_0$ related to the intensity of the spin-orbit interaction.
The absence of $\exp{-2qb}$ in the expression for $j_H$ further indicates that this predicted Hall and spin
Hall effects have a surface nature in contrast to the tunnelling current.

Up to now, the case of ``skew'' scattering was considered. In addition to this scattering mechanism, another
contribution to Hall and spin Hall currents originates from another term in the operator
of quantum mechanical velocity, proportional to spin-orbit interaction:
\begin{equation}
\hat v=\frac{d}{dt}\vec r=-i[\vec r\times\vec H]=\frac{\hbar\vec k}{m}
+\lambda[\vec\sigma\times\vec\nabla V(\vec r)],
\end{equation}
where $V(\vec r)$ -- potential of impurity, $\lambda$ -- spin-orbit constant. This additional contribution
to the Hall current is equivalent to a ``side jump'' mechanism~\cite{andp.200510206}.
In the present case it is written in final form as:
\begin{equation}
\label{eq:j_p_h_sj}\nonumber
\left<j_{l+r}^{\uparrow+\downarrow}\right>_{\mathrm{AHE}}^{\mathrm{P, sj}}=
\frac{2}{3\pi}\,\frac{e^2}{2\pi\hbar}\,\frac{\tilde \lambda c}{Ub}
\left(E_F^\uparrow k_F^\uparrow-E_F^\downarrow k_F^\downarrow\right)V,
\end{equation}
\begin{equation}
\label{eq:j_p_sh_sj}\nonumber
\left<j_{l+r}^{\uparrow+\downarrow}\right>_{\mathrm{SHE}}^{\mathrm{P, sj}}=
\frac{2}{3\pi}\,\frac{e^2}{2\pi\hbar}\,\frac{\tilde \lambda c}{Ub}
\left(E_F^\uparrow k_F^\uparrow+E_F^\downarrow k_F^\downarrow\right)V,
\end{equation}
\begin{multline}
\label{eq:j_ap_h_sj}\nonumber
\left<j_{l+r}^{\uparrow+\downarrow}\right>_{\mathrm{AHE}}^{\mathrm{AP, sj}}=
\frac{4}{15\pi}\,\frac{e}{2\pi\hbar}\,\frac{\tilde \lambda c}{Ub}
\Bigl(E_F^{\uparrow 2} k_F^\uparrow-E_F^{\downarrow 2} k_F^\downarrow\\
+\left(E_F^{\uparrow 2}+eV\right)^2\sqrt{k_F^{\uparrow 2}+eV}
-\left(E_F^{\downarrow 2}+eV\right)^2\sqrt{k_F^{\downarrow 2}+eV}
\Bigr),
\end{multline}
\begin{equation}
\label{eq:j_ap_sh_sj}\nonumber
\left<j_{l+r}^{\uparrow+\downarrow}\right>_{\mathrm{SHE}}^{\mathrm{AP, sj}}=
\left<j_{l+r}^{\uparrow+\downarrow}\right>_{\mathrm{SHE}}^{\mathrm{P, sj}}
\end{equation}

First of all, we note that both contributions into the Hall and spin Hall
currents are proportional to the concentration of impurities. This contrasts
to the usual Hall conductivity in ferromagnetic metals which is inversely
proportional to this concentration for the skew scattering and does not depend
on concentration for the side jump mechanism. However in the present case, Hall
current in metallic ferromagnetic electrodes is proportional to the current in this
electrode, itself proportional to the small parameter $\exp{-2qb}$. Therefore,
for thick enough insulating spacer, Hall and spin Hall effects inside the spacer
may become much larger than the corresponding effects within the ferromagnetic electrodes.

To find the Hall voltage $V_H$, we divided the expressions for Hall current by
conductance in $y$-direction:
\begin{multline}
G=\frac{e^2}{2\pi\hbar}\frac1b\sqrt{\frac{2m}{\hbar^2}U}\\
\times\left[
\left(1-\sqrt{1-\frac{E_F^\uparrow}{U}}\right)
+\left(1-\sqrt{1-\frac{E_F^\downarrow}{U}}\right)
\right].
\end{multline}

Estimated value of $V_H$ is $(10^{-5}$ to $10^{-3})V$, for $\tilde\lambda$
in interval $(10^{-2}$ to $10^{-1})$ and $c$ in interval $(0.01$ to $0.1)$. The most
interesting conclusion is the existence of a Hall voltage for AP-configuration
even in absence of any applied voltage. It means that an insulating spacer sandwiched
between two ferromagnetic electrodes in AP-configuration exhibits a spontaneous electric
polarization i.e. a spontaneous transverse ferroelectricity due to proximity effect.
The latter results from the asymmetric scattering on spin-orbit impurities of tunnelling
electrons penetrating into the insulating barrier from the ferromagnetic electrodes.

To experimentally measure this effect, one possibility would be to make electrically isolated metallic
islands aside of the tunnel barrier. These islands would get charged by electrostatic influence
with the charges arising on the side walls of the tunnel barrier. Measuring the voltage between
these islands and the MTJ electrodes in parallel and antiparallel magnetic configuration with an
electrometer could allow to detect and measure this new phenomenon of spontaneous transverse
ferroelectricity in MTJ.

\textbf{Acknowledgements:} This work was partly funded by the European Commission
through the ERC HYMAGINE grant n$^{\mathrm{o}}$ ERC 246942.

\end{document}